\documentclass[%
 preprint, 
 amsmath,amssymb,
 aps, physrev,
]{revtex4-2}

\usepackage{graphicx}
\usepackage{dcolumn}
\usepackage{bm}
\usepackage{booktabs}

\begin{document}

\preprint{APS/123-QED}

\title{\textbf{Turbulence teaches equivariance to neural networks} 
}%

\author{Ryley McConkey}
 \email{Contact author: rmcconke@mit.edu}
 
\author{Julia Balla}%

\author{Jeremiah Bailey}

\author{Ali Backour}

\author{Elyssa Hofgard}

\author{Tommi Jaakkola}

\author{Abigail Bodner}

\author{Tess Smidt}

\affiliation{%
  Massachusetts Institute of Technology\\
  Cambridge, Massachusetts, United States
}%

\date{\today}

\begin{abstract}
We show that the rotational nature of turbulence affects how neural networks learn mappings between quantities governed by the Navier-Stokes equations. We train super-resolution models at different wall-normal locations in a turbulent channel flow, where anisotropy varies naturally, and test their generalization to new coordinate frames, new anisotropy regimes, and a higher Reynolds number. Our findings inform both the design of equivariant machine learning models for turbulence and our understanding of how turbulence shapes what those models learn. First, mappings that better respect the rotational symmetries of the Navier-Stokes equations generalize better to new flows. Coordinate-frame generalization is therefore a key part of the broader generalization problem, since turbulent flows contain a wide range of local orientations. Second, turbulence itself partially teaches equivariance to learned mappings, an effect we call \textit{implicit data augmentation}. The effect strengthens with dataset size and with isotropy, since a more isotropic dataset samples more orientations under which the Navier-Stokes equations are covariant. Implicit augmentation is also scale-dependent, with smaller scales exhibiting lower equivariance error. This scale-dependency is consistent with Kolmogorov's hypothesis of local isotropy. Third, enforcing equivariance as an architectural inductive bias is the limit of these effects: an exactly equivariant network outperforms unconstrained CNNs on all generalization tests, with roughly an order of magnitude fewer parameters. We expect these effects to apply broadly to learned mappings between tensorial flow quantities, making them relevant to most machine learning applications in turbulence.

\end{abstract}

\maketitle


\section{Introduction}\label{sec:intro}

Turbulence is characterized by coherent, rotating eddies across a wide range of scales. An ensemble of turbulence realizations defines a probability distribution over flow states, and this probability distribution can itself possess symmetries. In isotropic turbulence, while any single realization lacks rotational symmetry, the ensemble is rotationally invariant. Kolmogorov's local isotropy hypothesis~\citep{Kolmogorov1991} extends this idea to bounded flows. At sufficiently high Reynolds number, the small-scale motions of any turbulent flow recover statistical isotropy, even when the large scales are anisotropic. The rotational symmetry of the Navier-Stokes equations is therefore reflected in the statistical structure of turbulence itself, most fully at the isotropic limit. 

Given the growing interest in data-driven emulators of the Navier-Stokes equations, this correspondence raises two questions. Can turbulence itself impart the symmetries of the underlying dynamics to learned mappings? Conversely, do models that better capture these symmetries generalize better to new distributions? Data-driven turbulence closure models, accelerated solvers, and super-resolution models are all motivated by the immense computational cost of simulating turbulence~\citep{Duraisamy2019}. Such approaches promise significant speedups, enabling progress in physics and engineering applications involving turbulence. However, the accuracy, generalizability, and trust placed in these models strongly depend on their physical consistency with the Navier-Stokes equations. Here, we ask whether the rotational structure of turbulence itself plays a role in making these emulators more physically consistent and better at generalizing.

Emulating the Navier-Stokes equations in a physically consistent way requires matching symmetries. The rotational covariance of the equations, reflected in some cases by the statistical structure of turbulence, should also be reflected in any learned mapping between flow quantities. \textit{Equivariance} is often an important property in machine learning for physics, as it ensures that models respect the symmetries of the underlying system. An equivariant model $f$ satisfies $f(g \cdot x) = g \cdot f(x)$, where $g$ is a given element of a symmetry group (such as a rotation), and $x$ is an input quantity. An equivariant model's outputs transform consistently with its inputs, matching the covariance of the physical system. Efforts have been made in nearly all domains of machine learning for turbulence to embed equivariance into machine learning models. For example, in turbulence closure modelling for Reynolds-averaged Navier-Stokes (RANS) problems, the commonly used Tensor Basis Neural Network (TBNN) achieves equivariance by predicting invariant coefficients of basis tensors that transform with $g$ \citep{Ling2016}. In large eddy simulation (LES), example strategies for embedding equivariance involve predicting invariant scalars \citep{kurz2025harnessingequivariancemodelingturbulence} or transforming into the eigenframe of the strain-rate tensor \citep{PRAKASH2022115457}. These approaches require task-specific architectural constructions, and each is tied to the particular flow quantity being modelled. It remains unclear how much equivariance comes from turbulence itself, and whether general-purpose equivariant architectures can perform well in turbulence.

Super-resolution (SR) with machine learning has become a promising method to augment numerical simulations of turbulence, by boosting the effective resolution of expensive calculations \citep{Duraisamy2019}. In this setting, convolutional neural networks (CNNs) have been widely applied to reconstruct velocity and vorticity fields from coarse inputs \citep{Fukami_Fukagata_Taira_2019, Liu2020, FukamiFukagataTaira_2024, Pang2024}. Complementary approaches span models that incorporate temporal coherence and dynamics-aware training objectives \citep{Fukami_Fukagata_Taira_2021, Page_2025}, as well as generative methods such as generative adversarial networks (GANs) \citep{Nista2024} or diffusion models that reproduce realistic spectra and scaling laws \citep{Shu2023,WhittakerNairLivescuChertkov_2024, Fan2025}. A rotationally equivariant super-resolution model would produce a super-resolved flow that transforms consistently when the input is rotated, matching the rotational covariance of the underlying physics. None of the SR methods cited above enforce this property, making super-resolution an ideal test-bed to investigate how symmetries emerge from turbulence data alone, and what changes when equivariance is enforced by construction.

Another way to encourage equivariance is via \textit{explicit} data augmentation. The objective of explicit data augmentation is to generate distributional symmetry, which is in turn imparted onto the mapping. Input/output pairs are randomly transformed during training, by sampling members of the desired symmetry group. In this work, we show that turbulence can achieve this effect \textit{implicitly}. In principle, a completely isotropic turbulence dataset knows no coordinate frame. In the infinite data limit, a model trained on a completely isotropic turbulence dataset should learn the rotational symmetries of the Navier-Stokes equations, having encountered similar structures in infinitely many orientations. Therefore, turbulence itself can produce an effect that often needs to be explicitly enforced in learned mappings. Equivariance can also be enforced exactly through architectural constraints, which results in data efficiency gains for atomistic systems~\citep{Batzner2022}. We explore equivariance as an inductive bias in this work.

In this study, we demonstrate three complementary results arising from the nature of turbulence. First, mappings which better respect the rotational symmetries of the Navier-Stokes equations generalize better to new flows. Second, the rotational nature of turbulence itself imparts rotational equivariance to learned mappings. Third, enforcing equivariance as an architectural inductive bias is the limit of these effects: our equivariant network outperforms the unconstrained CNN and its augmented counterpart on three generalization tests, while using roughly an order of magnitude fewer parameters.

The remainder of this paper is organized as follows. Section~\ref{sec:methodology} describes the methodology, including the equivariance error definition, the turbulent channel flow dataset, the CNN and equivariant CNN architectures, and the explicit data augmentation procedure. Section~\ref{sec:results} presents the results: Section~\ref{sec:generalization} demonstrates the correlation between equivariance error and generalization error, and Section~\ref{sec:equiv_error_N} demonstrates the implicit data augmentation and its scale-dependence. Section~\ref{sec:conclusion} concludes.

\section{Methodology}\label{sec:methodology}
\begin{figure}
    \centering
    \includegraphics[width=\textwidth]{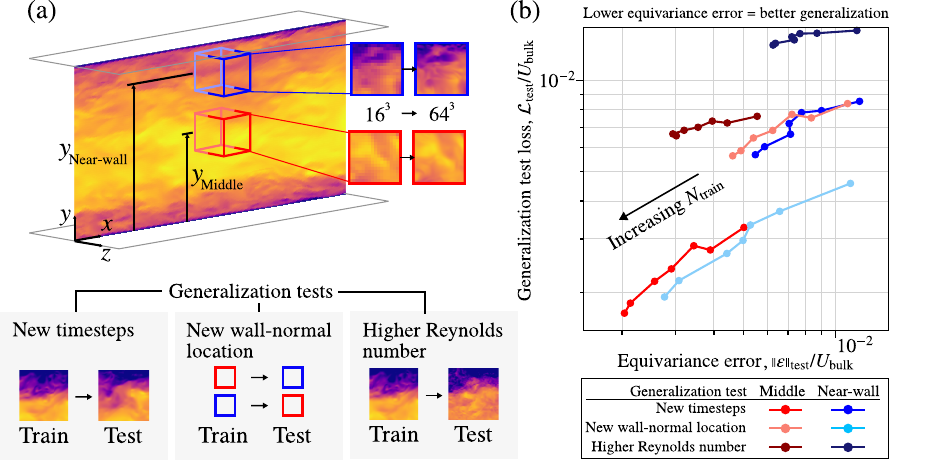}
    \caption{(a) Locations of the two sub-boxes (near-wall and middle) in the turbulent channel flow. (b) Test loss against equivariance error for three generalization tests on CNNs trained with explicit augmentation: held-out time steps, a change in wall-normal anisotropy regime, and a $5\times$ increase in Reynolds number. Colour denotes the training dataset. Lower equivariance error correlates with lower generalization error across all configurations.}
    \label{fig:boxes_generalization}
\end{figure}

The methodology is organized as follows. Section~\ref{sec:equivariance_error} defines the equivariance error, Section~\ref{sec:symmetry_group} discusses our choice of symmetry group, and the remainder of Section~\ref{sec:methodology} describes the dataset and model architectures. Figure~\ref{fig:boxes_generalization} shows the dataset configuration and the key finding from Section~\ref{sec:generalization}.

\subsection{Equivariance error}\label{sec:equivariance_error}

Before defining equivariance error, it is worth distinguishing two related but distinct notions of symmetry. A particular solution of the Navier-Stokes equations may or may not be symmetric under a given group action. A turbulent channel flow solution is not completely rotationally symmetric because the walls and the streamwise pressure gradient pick out preferred directions. A single realization of homogeneous isotropic turbulence is also not rotationally symmetric, even though the ensemble statistics are. Ensemble (statistical) symmetry is a property of a particular flow.

Equivariance is a property of the mapping. The Navier-Stokes equations themselves are covariant under rotations: if $\mathbf{u}(\mathbf{x},t)$ is a solution, then so is $R\mathbf{u}(R^{-1}\mathbf{x},t)$ for any rotation $R$, with the pressure and forcing transformed accordingly. Any mapping between flow quantities that arises from these equations should inherit this covariance, regardless of whether the specific solutions sampled by the training data happen to be symmetric. A super-resolution model, a closure model, or an accelerated solver should all transform their outputs consistently when their inputs are rotated, because the underlying physics does.

This distinction matters for two reasons. First, it clarifies what we are measuring. Equivariance error quantifies how well a learned mapping respects the covariance of the governing equations, not how symmetric the training data is. Second, it explains why equivariance is a useful property even for flows whose solutions break the symmetry. A near-wall channel flow has no rotational symmetry as a solution, but a model that maps coarse near-wall velocity fields to fine ones should still respect the rotational covariance of the equations that produced those fields. If it does not, it is encoding a coordinate-frame dependence that the physics does not have.

We quantify how well a model has learned equivariance via the \textit{equivariance error}~\citep{wang2022approximatelyequivariantnetworksimperfectly}. A perfectly equivariant model automatically transforms its outputs when the inputs transform. Formally, given input $x$, group $G$, and group elements $g\in G$, a model $f$ is $G$-equivariant if $f(g\cdot x) = g\cdot f(x) \ \forall  \ g \in G$. We define the pointwise equivariance error as the absolute residual of the definition of equivariance: 
\begin{align}
        \mathbf{E} (\mathbf{x}_n;g) &= \left|\mathbf{f}(g\cdot \mathbf{x}_n) - g\cdot \mathbf{f}(\mathbf{x}_n)\right|  
\end{align}
where $\mathbf{E}(\mathbf{x}_n;g)$ is the equivariance error for a given group element $g$ and coarsened velocity vector field $\mathbf{x}_n$, $n$ indexes the dataset, and $\mathbf{f}$ is the learned super-resolution function (a vector-valued function).

In this context, $\mathbf{E}(\mathbf{x}_n;g)$ is the component-wise absolute difference between two vector fields and is therefore a sign-fixed vector field. At each field point, it measures how close the mapping $\mathbf{f}$ is to being perfectly equivariant under a given group element $g$ for the input $\mathbf{x}$. We can collapse this vector field to generate integral quantities by taking the average over all group elements, data points, and field points. The average equivariance error norm is calculated as 
\begin{equation}
||\varepsilon|| = \frac{1}{|G|N} \sum_{g\in G} \sum_{n=1}^{N} || \mathbf{E}(\mathbf{x}_n;g)|| \,  \label{eq:equiv_error_norm} 
\end{equation}
where $N$ is the dataset size, $|G|$ is the cardinality of the group, and $||\cdot||$ is the $L^2$ norm. $||\varepsilon||$ is a single measure of how perfectly equivariant a vector-valued model is.

It should be noted that equivariance error does not depend at all on \textit{how well} a model performs with respect to the ground truth data---it only depends on how equivariant the model is. Therefore, it can be measured for any model without ground truth data. In this investigation, we show that equivariance error correlates with performance on ground truth data due to the nature of turbulence.

\subsection{Symmetry group}\label{sec:symmetry_group}
We focus in this work on the rotational octahedral group $O$. As discussed in Section~\ref{sec:equivariance_error}, our goal is to study how well a learned mapping respects the rotational symmetries of the Navier-Stokes equations, which are covariant under the full rotation group $SO(3)$~\citep{Pope_2000}. The channel walls and the streamwise pressure gradient break this symmetry at the level of any individual solution, but we do not claim the channel flow itself is $SO(3)$-symmetric. We ask whether a learned mapping between flow quantities respects the rotational covariance of the underlying equations.

We use a discrete subgroup of $SO(3)$ rather than the full continuous group for two practical reasons. First, the underlying numerical simulation only embeds the discrete symmetries of the equations, since discretization error depends on the alignment of the flow with the grid~\citep{AgdesteinVerstappenSanderse_2026}. Recent work in LES closure modelling has further emphasized that enforcing continuous symmetries in the discrete equations can be unnecessary or even harmful~\citep{AgdesteinSanderse_2026}. Second, augmenting with continuous rotations requires interpolating fields onto a transformed grid, which introduces an additional source of error~\citep{YasudaOnishi_2023}. The 24 rotations in $O$ permute voxels exactly onto other voxels on a uniform Cartesian grid, so applying any $g \in O$ introduces no interpolation error.
We restrict our attention to rotations rather than the full octahedral symmetry group (which also includes reflections and inversions), because the present investigation concerns the rotational nature of turbulence specifically. We leave reflections and inversions to future work. With $G = O$ in Equation~\ref{eq:equiv_error_norm}, $|G| = 24$, and $g \in O$ are the 24 octahedral rotations.

\subsection{Dataset and model details}

For a representative turbulence dataset, we select the canonical turbulent channel flow. In this flow, rotating eddies of various scales pass through the domain and are influenced anisotropically by the presence of the top and bottom walls. The degree of anisotropy varies with the wall-normal direction. Near the wall, flow is highly anisotropic. In the center of the channel, the flow is still anisotropic but to a much lower degree. Choosing turbulent channel flow allows us to vary the degree of isotropy in the training dataset by sub-sampling the domain. However, we are still able to make direct comparisons between different sub-sampled regions of the flow, since the fields come from the same flow. Specifically, we use the Johns Hopkins turbulent channel flow dataset at friction Reynolds number $Re_\tau = 1000$ \citep{Li_2008}. This moderately turbulent Reynolds number produces rotating eddies of various sizes in the channel. In Section~\ref{sec:generalization}, we test generalization up to $Re_\tau = 5200$ from the same database \citep{Li_2008}.

\begin{figure}
    \centering
    \includegraphics[]{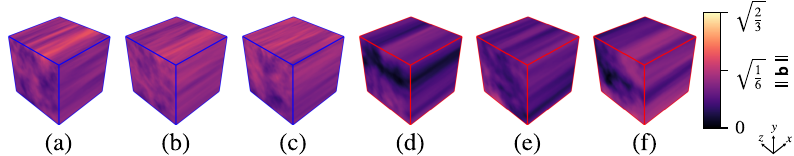}
    \caption{Anisotropy tensor (Eq.~\ref{eq:anisotropy_tensor_defn}) magnitude for the three sub-boxes in the near-wall (blue) and middle (red) locations in the channel. The $\sqrt{2/3}$ bound represents the one-component limit \citep{Banerjee2007}. The near-wall region is more anisotropic due to the proximity of the wall.}
    \label{fig:anisotropy_contours}
\end{figure}

In two distinct regions of the flow, we take three sub-boxes each, creating an ensemble of six boxes total. The two regions correspond to the middle of the channel and the near-wall region. Figure~\ref{fig:boxes_generalization} shows the locations of the two regions, and an example super-resolution task. Within each region, the three sub-boxes are statistically equivalent, allowing us to test the effect of spatio-temporal ensembling on the model's behaviour in Section~\ref{sec:results}. Figure~\ref{fig:anisotropy_contours} shows how the degree of anisotropy varies between the two selected $y$ coordinates by comparing the magnitude of the anisotropy tensor $\mathbf{b}$.  The anisotropy tensor is calculated by
\begin{equation}
    \mathbf{b} = \frac{\langle \mathbf{U}' \otimes \mathbf{U}' \rangle}{2k} - \frac{1}{3}\mathbf{I}
    \label{eq:anisotropy_tensor_defn}
\end{equation}
where $\mathbf{U}' = \mathbf{U} - \langle \mathbf{U} \rangle$ is the fluctuating velocity and $k = \frac{1}{2}\langle \mathbf{U}' \cdot \mathbf{U}' \rangle$ is the turbulent kinetic energy. The magnitude of $\mathbf{b}$ corresponds to the degree of anisotropy. Figure~\ref{fig:anisotropy_contours} shows that the near-wall region is clearly more anisotropic than the mid-channel region, given the higher magnitude of the anisotropy tensor.

All three boxes from a distinct region have the same central $y$ coordinate. Given that turbulent channel flow is statistically stationary and homogeneous at a fixed $y$, additional samples in time and additional boxes (at fixed $y$) simply add data from a given ensemble of identically distributed turbulence realizations. 
 
The dataset used consists of a variety of turbulent channel flow ensembles. The aim of the dataset is to explore various methods of ensembling turbulence realizations. We generate these ensembles using both a single time series (temporal ensembling), and combining multiple times series from multiple sub-boxes (spatio-temporal ensembling). We split the Johns Hopkins channel flow dataset~\citep{Li_2008} of 4000 snapshots over approximately one flow-through time sequentially into a training set (67\%), validation set (20\%), and test set (13\%). We sub-sample six square boxes of side length $0.6H$, three of which have centers at $y_\mathrm{Near\text{-}wall} = 1.65H$ and three of which have centers at $y=H$, where $H$ is the channel half-height. We randomly choose $x$ and $z$ coordinates in the channel for the three boxes. We explore randomly selected subsets of the training dataset in Section~\ref{sec:results}, including randomly sampling the time series from a single-box dataset (temporal ensembling), and a three-box dataset (spatio-temporal ensembling). 

All code used in this study is available on GitHub: \url{https://github.com/atomicarchitects/turbulence-implicit-augmentation} \citep{code_github}. We include end-to-end steps to reproduce our results, model weights, and all code used to generate the plots in this manuscript.

\subsubsection{Convolutional neural network (CNN) model}\label{sec:cnn_model}
The goal of this work is not to develop a state-of-the-art super-resolution model, but to investigate the relationship between equivariance error and generalization performance. We therefore select a simple and representative super-resolution architecture as a baseline~\citep{dong2016image}. Specifically, we use a compact multi-scale convolutional super-resolution network that upsamples a low-resolution velocity field volume $\mathbf{x} \in \mathbb{R}^{3\times D \times H \times W}$ to the target high-resolution $\mathbf{U} \in \mathbb{R}^{3\times sD \times sH \times sW}$. Upsampling by factor $s$ is implemented as a sequence of resize-then-refine stages, one for each factor of 2. At each stage, the input is upsampled through trilinear interpolation and passed through two convolutional layers. A final convolution projects the features to 3 output channels, yielding the super-resolved prediction. We use reflection padding on all image edges. The SR model is trained to minimize the mean absolute error (MAE) loss between the ground truth and predicted high-resolution fields, which has been shown to better preserve perceptual quality and reduce oversmoothing compared to mean squared error (MSE) loss in image restoration tasks \citep{Zhao2017}.

We fix the CNN architecture and hyperparameters for all of our experiments. The network contains two successive upsampling layers, each enlarging the input by a factor of $2$, resulting in an overall scale factor of $s=4$. 3D Convolutions use kernels of size $3$ with reflection padding of one pixel on each side, followed by ReLU activations. All hidden layers have 128 channels. Models are trained with the Adam optimizer with learning rate $6\times10^{-4}$ and batch size of $32$. We train each model for 3000 epochs and take the model which has the lowest validation loss over all epochs.  The required epochs for training convergence increased with the number of training points. However, we observed that the training converged for all models and datasets within the 3000 epoch range.

\subsubsection{Equivariant CNN model}\label{sec:escnn_model}

We compare the CNN baseline against an architecture that enforces equivariance to the rotational octahedral group $O$ as an inductive bias. Equivariant convolutional architectures replace standard convolutions with group convolutions, in which the same kernel is applied at each orientation of the symmetry group rather than at a single fixed orientation~\citep{e2cnn,cesa2022a}. The kernel weights are shared across all 24 elements of $O$, so the resulting layer transforms outputs consistently when inputs are rotated by any $g \in O$. We use the ESCNN library~\citep{cesa2022a} to construct an equivariant super-resolution network (srESCNN) that mirrors the CNN architecture as closely as possible.

The srESCNN takes a vector field input and produces a vector field output, both in the standard 3D vector representation of $O$. Hidden layers use the regular representation, which is the most expressive choice (each hidden feature carries a value for each of the 24 group elements) and corresponds to standard group convolution. We use 16 copies of the regular representation in the hidden layers, giving an effective hidden width of $16 \times 24 = 384$ features. The architecture matches the CNN: an input convolution, two upsampling stages (each consisting of a trilinear upsample followed by two convolutional layers with ReLU activations), and a final $1\times 1$ convolution projecting to the output. All convolutions use kernel size 3 with reflection padding. Despite the wider hidden representation, weight sharing across $O$ reduces the total parameter count to 246{,}368, compared to 1{,}791{,}238 for the CNN.

The srESCNN is trained with the same loss, optimizer, learning rate, batch size, and number of epochs as the CNN baseline (Section~\ref{sec:cnn_model}). The srESCNN trains in comparable wall-clock time to the CNN despite having fewer parameters, reflecting the higher per-operation cost of equivariant layers. We verify equivariance numerically: applying any $g \in O$ to the input and comparing $g \cdot \mathbf{f}(\mathbf{x})$ to $\mathbf{f}(g \cdot \mathbf{x})$ yields differences at the level of floating-point accumulation error ($\sim 10^{-5}$).

\begin{figure}
    \centering
    \includegraphics[width=\textwidth]{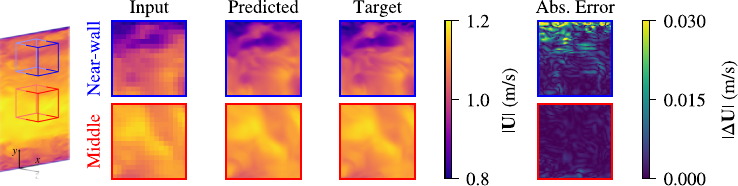}
    \caption{Example outputs from the CNN (1500 samples, single box, rotational augmentation). Colour here is by velocity magnitude, but the models super-resolve each velocity component separately.}
    \label{fig:example_predictions}
\end{figure}
\subsubsection{Explicit data augmentation}

Explicit data augmentation is a standard technique for encouraging equivariance in learned mappings by directly enforcing distributional symmetry on the training data~\citep{wang2022approximatelyequivariantnetworksimperfectly}. During training, input/output pairs are transformed by group elements drawn at random from the desired symmetry group, exposing the model to the same underlying structures across many orientations. In the limit of infinite samples, this procedure produces a training distribution that is invariant under the action of $G$, which in turn encourages the learned mapping $\mathbf{f}$ to satisfy $\mathbf{f}(g\cdot \mathbf{x}) = g\cdot \mathbf{f}(\mathbf{x})$ for all $g\in G$. In this work, we use explicit augmentation as a baseline against which to compare the implicit augmentation arising from turbulence itself.

In explicit data augmentation, at each training iteration we draw a group element $g$ uniformly at random from the rotational octahedral group $O$ and apply it jointly to the low-resolution input and the high-resolution target. Let $\mathbf{u}[\mathbf{i}] \in \mathbb{R}^3$ denote a discrete velocity field defined on a uniform Cartesian grid, where $\mathbf{i} \in \mathbb{Z}^3$ indexes voxels relative to the grid center, and let $R_g$ be the rotation matrix representing $g \in O$. The group action is then
\begin{equation}
    (g\cdot \mathbf{u})[\mathbf{i}] = R_g \, \mathbf{u}\!\left[R_g^{-1}\mathbf{i}\right],
    \label{eq:vector_field_action}
\end{equation}
in which the voxel index is pulled back by $R_g^{-1}$ while the vector value at each voxel is rotated by $R_g$. For an input--target pair $(\mathbf{x}_n, \mathbf{U}_n)$, the augmented pair $(\tilde{\mathbf{x}}_n, \tilde{\mathbf{U}}_n)$ is then
\begin{equation}
    \tilde{\mathbf{x}}_n = g\cdot \mathbf{x}_n, \qquad
    \tilde{\mathbf{U}}_n = g\cdot \mathbf{U}_n, \qquad
    g \sim \mathrm{Uniform}(O),
    \label{eq:explicit_augmentation}
\end{equation}
with $g\cdot$ acting on each field according to~(\ref{eq:vector_field_action}). Because $|O|=24$ and our grid is uniform, all 24 rotations permute voxels exactly onto other voxels, so~(\ref{eq:explicit_augmentation}) introduces no interpolation error. We sample an independent $g$ for every training example at each epoch, ensuring that over the course of training the model sees each example in many distinct orientations. For further details, we refer the reader to the source code~\citep{code_github}.

\section{Results}\label{sec:results}

For a discussion of the distinction between \textit{explicit} and \textit{implicit} data augmentation and the rotational augmentation group used in this work, see Section~\ref{sec:symmetry_group}. In Section~\ref{sec:results}, ``augmentation'' refers to explicit augmentation---we show that implicit augmentation occurs automatically. Specifically, when a training point is sampled from the original dataset, input/output pairs are randomly transformed with some group element $g\in O$ (see Section~\ref{sec:equivariance_error} for more details).

Figure~\ref{fig:example_predictions} shows a super-resolved test set example using a rotationally augmented model. Both the augmented and non-augmented models extrapolate well in time (in the original coordinate frame) after seeing 1500 examples from a given box.

\begin{figure}
    \centering
    \includegraphics[]{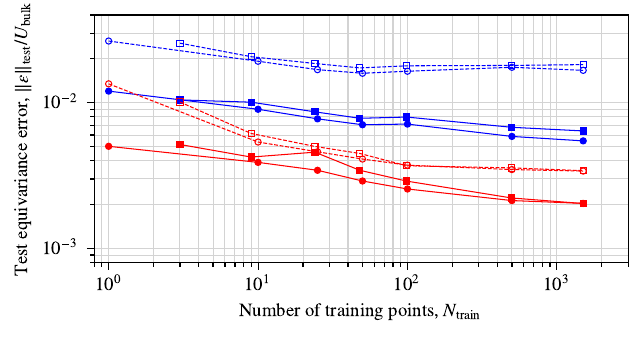}
    \caption{Test equivariance error against the number of training samples for CNNs trained on the near-wall and middle datasets, with and without explicit data augmentation. Equivariance error decreases with the number of training samples in every case. The effect is stronger for the mid-channel data than for the near-wall data.}
    \label{fig:equiverror_vs_samples}
\end{figure}

\subsection{Equivariance and generalization to new flows}\label{sec:generalization}
Models in machine learning for turbulence frequently struggle to generalize to new geometries, flow conditions, and Reynolds numbers \citep{Duraisamy2019}. The causes include insufficient training data, distributional shift from subtle changes in physics, and emergent phenomena not represented in the training dataset. We ask whether a model that better captures the rotational symmetries of the Navier-Stokes equations also generalizes better.

Figure~\ref{fig:boxes_generalization}(b) compares generalization losses to equivariance error magnitude for three tests: extrapolating in time, generalizing to a new anisotropy regime by changing the wall-normal location, and generalizing to a $5\times$ increase in Reynolds number. Across all three tests, lower equivariance error correlates with lower generalization error. The CNN trained on the more isotropic mid-channel data has lower equivariance error and generalizes better than the CNN trained near the wall. Explicit data augmentation reduces equivariance error and improves generalization for both training datasets. The model trained on the near-wall dataset generalizes better to the middle of the channel than the reverse case, likely because the more isotropic middle distribution does not generalize well to the strongly anisotropic flow near the wall. All models feature higher test loss when generalizing to $Re_\tau = 5200$, since they have not seen the broader inertial range and higher levels of turbulence in that test dataset. Even for this challenging test, however, the correlation between equivariance error and generalization error holds for almost all models. Note that the $\|\varepsilon_\mathrm{test}\|$ reported in Figure~\ref{fig:boxes_generalization}(b) is the test set equivariance error from the original time series for all three plots, so this is the correlation between equivariance error on a model's original test dataset (held out time steps) and how well that model generalizes to a new turbulent flow.

If equivariance error and generalization error are correlated, what happens at the limit of zero equivariance error? The srESCNN architecture (Section~\ref{sec:escnn_model}) enforces equivariance to $O$ as an inductive bias, achieving equivariance error at the level of floating-point accumulation ($\sim 10^{-5}$). Table~\ref{tab:generalization} compares the srESCNN against the CNN baselines on all three generalization conditions. The srESCNN achieves the lowest test loss and lowest equivariance error in every column, despite using roughly an order of magnitude fewer parameters than the CNN. The improvement over CNN(aug) is modest in the same-flow tests, where both models perform similarly on held-out time steps, but grows under distribution shift. This is the expected pattern if equivariance error is what matters for generalization in turbulence. When the test distribution overlaps the training distribution, the loss benefit of zero equivariance error is small, since the model rarely encounters orientations it has not effectively seen. When the test distribution shifts (new Reynolds number, new anisotropy regime), the model is forced to handle structures in orientations the training data did not cover, and equivariance is an advantage.

The srESCNN result also rules out an alternative reading of Figure~\ref{fig:boxes_generalization}(b): that the apparent correlation between equivariance error and generalization error is just a quirk of explicit data augmentation. One could argue that augmentation independently improves generalization for reasons unrelated to equivariance, and that the two effects only happen to be correlated because they both follow from augmenting the data. The srESCNN reduces equivariance error through a completely different mechanism, namely an architectural constraint with no augmentation, and it lands on the same trend. The correlation is therefore about equivariance error itself, not about augmentation as a technique.

\begin{table}
  \centering
  \caption{Test loss and equivariance error $\|\varepsilon\|$ for three generalization conditions, all at 1500 training samples from a single box. Same-flow: held-out time steps at $Re_\tau{=}1000$, evaluated on the same anisotropy regime as training. Reynolds generalization: $Re_\tau{=}5200$, same anisotropy regime. Anisotropy generalization: $Re_\tau{=}1000$, opposite anisotropy regime. Best result in each column is in bold. Loss is mean absolute error; the bulk velocity is unity in this dataset.}
  \label{tab:generalization}
  \footnotesize
  \begin{tabular}{llcccccc}
    \toprule
    & & \multicolumn{2}{c}{Same-flow} & \multicolumn{2}{c}{Reynolds gen.} & \multicolumn{2}{c}{Anisotropy gen.} \\
    \cmidrule(lr){3-4} \cmidrule(lr){5-6} \cmidrule(lr){7-8}
    Architecture & Params & Loss & $\|\varepsilon\|$ & Loss & $\|\varepsilon\|$ & Loss & $\|\varepsilon\|$ \\
    \midrule
    \multicolumn{8}{l}{\textit{Middle-trained}} \\
    CNN        & 1.79M & $1.9{\times}10^{-3}$ & $3.4{\times}10^{-3}$ & $6.7{\times}10^{-3}$ & $4.7{\times}10^{-3}$ & $6.1{\times}10^{-3}$ & $8.0{\times}10^{-3}$ \\
    CNN (aug)  & 1.79M & $1.7{\times}10^{-3}$ & $2.0{\times}10^{-3}$ & $6.6{\times}10^{-3}$ & $3.0{\times}10^{-3}$ & $5.6{\times}10^{-3}$ & $4.6{\times}10^{-3}$ \\
    srESCNN    & 246k  & $\mathbf{1.6{\times}10^{-3}}$ & $\mathbf{1.5{\times}10^{-5}}$ & $\mathbf{6.5{\times}10^{-3}}$ & $\mathbf{1.6{\times}10^{-5}}$ & $\mathbf{5.5{\times}10^{-3}}$ & $\mathbf{3.3{\times}10^{-5}}$ \\
    \midrule
    \multicolumn{8}{l}{\textit{Nearwall-trained}} \\
    CNN        & 1.79M & $5.7{\times}10^{-3}$ & $1.7{\times}10^{-2}$ & $1.3{\times}10^{-2}$ & $1.7{\times}10^{-2}$ & $2.1{\times}10^{-3}$ & $8.5{\times}10^{-3}$ \\
    CNN (aug)  & 1.79M & $5.7{\times}10^{-3}$ & $5.5{\times}10^{-3}$ & $1.3{\times}10^{-2}$ & $6.3{\times}10^{-3}$ & $1.9{\times}10^{-3}$ & $2.8{\times}10^{-3}$ \\
    srESCNN    & 246k  & $\mathbf{5.5{\times}10^{-3}}$ & $\mathbf{5.1{\times}10^{-5}}$ & $\mathbf{1.3{\times}10^{-2}}$ & $\mathbf{5.1{\times}10^{-5}}$ & $\mathbf{1.8{\times}10^{-3}}$ & $\mathbf{3.8{\times}10^{-5}}$ \\
    \bottomrule
  \end{tabular}
\end{table}

\subsection{Implicit spatio-temporal data augmentation in turbulence}\label{sec:equiv_error_N}
\subsubsection{Effect of increasing training data}

\begin{figure}
    \centering
    \includegraphics[width=\textwidth]{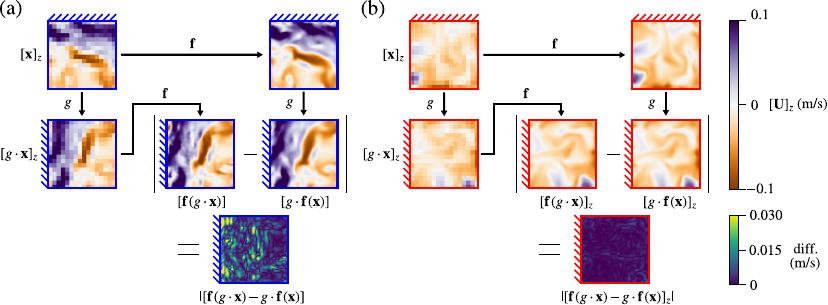}
    \caption{Example test set equivariance error field for the (a) near-wall and (b) middle datasets, and the non-augmented CNN model (Sec~\ref{sec:cnn_model}). We show the case of 1500 training examples from a single box, without any explicit data augmentation (i.e., implicit augmentation only). Hatching indicates the direction of the top wall for reference (not the wall boundary). $[\cdot]_z$ denotes the $z$-component of a field. The $z$ velocity component on an $xy$ plane is shown, with $g \equiv C_4z$ (a 90$^\circ$ rotation about the z axis). The $z$ velocity component is not transformed by this rotation, so the transformed image appears as a straightforward rotation. Comparing the difference contours, we see that the model trained on the more isotropic mid-plane data (b) has a significantly lower equivariance error. 
    }
    \label{fig:example_equivariance}
\end{figure}

We show that \textit{implicit} spatio-temporal data augmentation occurs in turbulence. With more examples of rotational turbulent flow, models learn equivariance without explicit data augmentation. Figure~\ref{fig:equiverror_vs_samples} shows the equivariance error against the number of training samples for models with and without explicit data augmentation. Equivariance error decreases with the number of training samples in every case. The decrease is faster for the more isotropic mid-channel data than for the near-wall data, which matches the intuition that a fully isotropic dataset contains complete distributional symmetry under rotation. The same effect appears spatially in Figure~\ref{fig:example_equivariance}. Without explicit augmentation, the model trained on the middle of the channel has substantially lower equivariance error than the near-wall model.

Even though implicit data augmentation reduces equivariance error, explicit data augmentation reduces it further (Figure~\ref{fig:equiverror_vs_samples}). The benefit of explicit augmentation is largest for the anisotropic near-wall data, where implicit augmentation provides the least coverage of the rotational symmetries. In the finite-data regime, with data that is not fully isotropic, explicit augmentation is still advantageous.

\subsubsection{Scale-dependence of learned equivariance}\label{sec:scale_dependence}
Kolmogorov's local isotropy hypothesis states that small-scale motions recover statistical isotropy when the Reynolds number is sufficiently high \citep{Kolmogorov1991}. In Section~\ref{sec:equiv_error_N}, we showed that a more isotropic dataset leads to lower equivariance error. Combined with Kolmogorov's hypothesis, this suggests that equivariance error itself should be scale-dependent: the largest scales of a turbulent channel flow are anisotropic, the smallest scales are closer to isotropic, and a model trained on this flow should reflect that variation in how well it learns equivariance at different scales.

\begin{figure}
    \centering
    \includegraphics[]{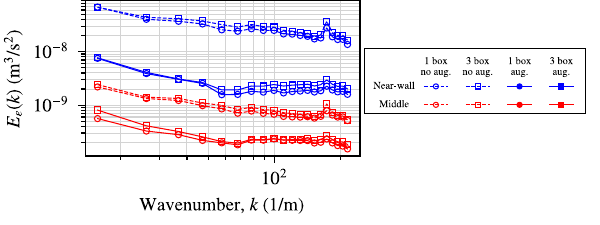}
    \caption{One-dimensional equivariance error power spectra for CNNs trained on the near-wall and middle datasets, with and without explicit data augmentation. Equivariance error decreases with wavenumber up to the input cutoff at $k\approx 90$, consistent with the scale-dependence of isotropy in channel flow. A sharp peak appears near $k\approx 180$, above the cutoff where the model is super-resolving in spectral space.}

    \label{fig:equiv_error_spectra}
\end{figure}

Figure~\ref{fig:equiv_error_spectra} shows the one-dimensional power spectra of the equivariance error vector. For all cases, the equivariance error decreases with wavenumber up to the input cutoff at $k\approx 90$. This trend is consistent with Kolmogorov's hypothesis of increasing isotropy at smaller scales, though the falloff alone is not a definitive test, since the underlying turbulent kinetic energy also decreases with $k$. Without explicit data augmentation, the falloff continues into the small scales. With explicit data augmentation, the trend flattens near the cutoff, since augmentation reduces equivariance error roughly equally across all resolved scales.

Above the input cutoff ($k > 90$), the model is super-resolving in spectral space. At these wavenumbers, neither implicit nor explicit augmentation can act, since augmentation only randomizes orientations of structures present in the input, and there is no input content above the cutoff to augment. We therefore expect higher equivariance error in this band, and Figure~\ref{fig:equiv_error_spectra} shows exactly that: a sharp peak appears near $k\approx 180$. The peak occurs at roughly the second harmonic of the cutoff wavenumber, which is consistent with the characteristic frequency response of the upsampling pipeline (trilinear interpolation followed by convolutional refinement at scale factor 2, applied twice).

These results confirm findings of the previous work by \citet{Wang2024}, which demonstrated that a symmetry-breaking model learns scale-dependent symmetry-breaking weights. Our results confirm this from a top-down viewpoint. While we show that a single model learns scale-dependent equivariance, \citet{Wang2024} showed that models trained on scale-dependent data learn different symmetry-breaking weights. Both investigations demonstrate that models learn scale-dependent equivariance, due to the scale-dependent nature of isotropy in a turbulent flow. The peak observed in Figure~\ref{fig:equiv_error_spectra} was not observed in \citet{Wang2024} because that work scale-separated the training data \textit{a priori}, while we train on multiple scales simultaneously.

\section{Conclusion}\label{sec:conclusion}

This paper makes three central claims about learned mappings between turbulence quantities that affect machine learning applications across turbulence:

\begin{enumerate}
    \item[(i)] Mappings that better respect the rotational symmetries of the Navier-Stokes equations generalize better to new flows. Across CNNs with and without explicit augmentation, lower equivariance error tracks lower generalization error on three distinct tests: held-out time steps, a $5\times$ Reynolds number increase, and a change in wall-normal anisotropy regime.
    \item[(ii)] Turbulence itself partially teaches equivariance to learned mappings, an effect we call \textit{implicit data augmentation}. We attribute this to the fundamentally rotational nature of turbulent flows. A turbulence realization presents structures in many orientations, partially covering the rotational symmetry of the Navier-Stokes equations. The effect strengthens with dataset size and with isotropy, since a more isotropic dataset samples more orientations under which the Navier-Stokes equations are covariant. Implicit augmentation is also scale-dependent, with smaller scales exhibiting lower equivariance error, consistent with Kolmogorov's hypothesis of local isotropy.
    \item[(iii)] Enforcing equivariance as an architectural inductive bias is the limit of (i) and (ii). Our srESCNN architecture, which is exactly equivariant to the rotational octahedral group, achieves the lowest test loss and lowest equivariance error in every generalization condition we tested, despite using roughly an order of magnitude fewer parameters than the CNN baseline.
\end{enumerate}

The correlation in (i) informs several future directions in the application of machine learning for turbulence. Though we selected super-resolution in this work, we expect (i) and (ii) to extend to other applications of machine learning in turbulence. We are now examining how implicit data augmentation occurs in subgrid scale modelling, where early results show similar behaviour. More broadly, we expect any learned mapping between quantities in a turbulent flow field to be subject to implicit data augmentation, since the underlying rotational nature of turbulence is the same. 

The field-wide problem of generalizability has sparked massive interest, given current unsatisfactory performance \citep{Duraisamy2019}. We demonstrate that a model's ability to generalize to new coordinate frames is correlated with its ability to generalize to new turbulent flows; improving coordinate-frame generalization therefore improves generalization to new flows directly. The present work also raises a concern with the use of large, statistically stationary, and isotropic datasets to validate new machine learning methodologies. Training under these ideal conditions promotes an overly optimistic view of new methodologies that does not transfer to anisotropic flows. 

This work has several limitations. We focus on channel flow at two wall-normal locations; while the conceptual argument that turbulence's rotational structure imparts equivariance is general, validating this on homogeneous isotropic turbulence and other flow geometries is a natural next step. The scale-dependence analysis in Figure~\ref{fig:equiv_error_spectra} reports the equivariance error spectrum directly. Future work will investigate the relationship between the equivariance error spectrum and the energy spectrum. The peak in the equivariance error spectrum near $k\approx 180$ matches what we expect from the upsampling pipeline's frequency response (trilinear interpolation followed by convolutional refinement at scale factor 2, applied twice), but understanding the precise mechanism requires additional study.

In general, the case for enforcing equivariance depends on the data regime. \citet{Brehmer2024} show that equivariance primarily improves data efficiency, and that non-equivariant models trained with data augmentation can close the gap given enough data and epochs. In the finite-data regime, the implicit augmentation in (ii) reduces equivariance error but does not eliminate it. Explicit data augmentation continues to reduce equivariance error on top of the implicit effect. The benefit of explicit augmentation is largest for anisotropic training data, where implicit augmentation provides less coverage of the rotational symmetries. We expect this finding to hold in the limit of abundant data, but most applications of machine learning in turbulence operate far from that limit. Direct numerical simulations are expensive, high-quality datasets are scarce, and practitioners routinely train models on a few thousand snapshots from a single flow configuration. In this regime, the inductive bias of an equivariant architecture is an advantage, since the data alone cannot fully cover the rotational symmetries of the Navier-Stokes equations. The implicit data augmentation in (ii) only partially closes that gap, and only when the training data is sufficiently isotropic.

Given this regime, we recommend enforcing equivariance as an architectural inductive bias. However, equivariant architectures require more careful implementation than standard CNNs and may not integrate easily with existing pipelines. If that effort is not feasible, we recommend using explicit data augmentation and including anisotropic flows in the training data. The srESCNN trains in comparable wall-clock time to the CNN, so the parameter savings come without runtime cost. For context, other architectural approaches can also impose exact equivariance, such as group-equivariant neural operators \citep{HelwigZhangFuEtAl_2023,XuHanLouEtAl_2024}, and regularization can encourage inexact equivariance without altering the backbone \citep{BaiEtAl_2025, Raissi2019}. The srESCNN result confirms what is known in several other scientific machine learning domains: architectures with equivariance as an inductive bias need less training data than non-equivariant ones. We argue that turbulence has an additional reason to favour equivariant architectures, beyond the standard data-efficiency argument. Because turbulence is fundamentally rotational, with eddies of every scale and orientation passing through any flow, generalizing to new coordinate frames and generalizing to new flows are tightly linked. New flows are hard to generalize to in part because they look like rotated versions of training flows. An equivariant architecture handles those rotations by construction. Combining an equivariant architecture with the implicit augmentation that turbulence provides is the most data-efficient path to learned mappings that respect the rotational symmetries of the Navier-Stokes equations.

\begin{acknowledgments}
We acknowledge the support of the National Science Foundation under Cooperative Agreement PHY-2019786 (The NSF AI Institute for Artificial Intelligence and Fundamental Interactions). J. Balla was supported by the Department of Defense (DoD) through the National Defense Science \& Engineering Graduate (NDSEG) Fellowship Program. J. Bailey was supported by the MIT Summer Research Program (MSRP). EH was supported by the U.S. Department of
Energy, Office of Science, Office of Advanced Scientific
Computing Research, Department of Energy Computational
Science Graduate Fellowship under Award Number DESC0024386. RM was supported by the Natural Sciences and Engineering Research Council of Canada (NSERC), the Thornton Family Fund, and the MIT Electrical Engineering and Computer Science Transformation Grant. This work was supported by the U.S. Department of Energy, National Nuclear Security Administration under Award Number DE-NA0004266. This research used resources of the MIT Office of Research Computing and Data.
\end{acknowledgments}


\bibliography{references}

\end{document}